\documentclass[a4paper,fleqn,usenatbib]{mnras}

\usepackage{newtxtext,newtxmath}

\usepackage[T1]{fontenc}
\usepackage{ae,aecompl}

\usepackage{graphicx}	
\usepackage{amsmath}	
\usepackage{amssymb}	

\newcommand{\um}{$\mu$m}

\title[MIR Polarimetry of AGN]{The origin of the mid-infrared nuclear polarization of active galactic nuclei}

\author[Lopez-Rodriguez et al.]{
E. Lopez-Rodriguez$^{1,2,3}$\thanks{E-mail: \href{mailto:elopezrodriguez@sofia.usra.edu}{elopezrodriguez@sofia.usra.edu}},
A. Alonso-Herrero$^{4,5,6}$,
T. Diaz-Santos$^{7}$, \newauthor 
O. Gonzalez-Martin$^{8}$, 
K. Ichikawa$^{9,6,10}$,
N. A. Levenson$^{11}$,
M. Martinez-Paredes$^{8}$,   \newauthor
R. Nikutta$^{12}$,
C. Packham$^{6,10}$,
E. Perlman$^{13}$,
C. Ramos Almeida$^{14,15}$,   \newauthor
J. M. Rodriguez-Espinosa$^{14,15}$,
C. M. Telesco$^{16}$
\\
	$^{1}$ SOFIA/USRA, SOFIA Science Center, NASA Ames Research Center, Moffett Field, CA 94035, USA \\
	$^{2}$Department of Astronomy, University of Texas at Austin, 1 University Station C1400, Austin, TX 78712, USA \\
	$^{3}$McDonald Observatory, University of Texas at Austin, Austin, TX 78712, USA \\
	$^{4}$Centro de Astrobiolog\'ia (CAB, CSIC-INTA), ESAC Campus, E-28692 Villanueva de la Ca\~nada, Madrid, Spain \\
	$^{5}$Department of Physics, University of Oxford, Oxford OXI 3RH, UK \\
	$^{6}$Department of Physics and Astronomy, University of Texas at San Antonio, One UTSA Circle, San Antonio, TX 78249, USA \\
	$^{7}$N\'ucleo de Astronom\'ia de la Facultad de Ingener\'ia, Universidad Diego Portales, Av. Ej\'ercito Libertador 441, Santiago, Chile \\
	$^{8}$Instituto de Radioastronom\'ia y Astrof\'isica (IRyA-UNAM), 3-72 (Xangari), 8701 Morelia, Mexico \\
	$^{9}$Department of Astronomy, Columbia University, 550 West 120th Street, New York, NY 10027, USA \\
	$^{10}$National Astronomical Observatory of Japan, 2-21-1 Osawa, Mitaka, Tokyo 181-8588, Japan \\
	$^{11}$Space Telescope Science Institute, 3700 San Martin Dr, Baltimore, MD 21218, USA \\
	$^{12}$National Optical Astronomy Observatory, 950 N Cherry Ave, Tucson, AZ 85719, USA \\
	$^{13}$Florida Institute of Technology, Melbourne, FL 32901, USA \\
	$^{14}$Instituto de Astrof\'sica de Canarias, C/V\'ia L\'actea, s/n, E-38205 La Laguna, Tenerife, Spain \\
	$^{15}$Departamento de Astrof\'isica, Universidad de La Laguna, E-38205 La Laguna, Tenerife, Spain \\
	$^{16}$Department of Astronomy, University of Florida, 211 Bryant Space Science Center, Gainesville, FL 32611-2055, USA
	 }

\date{Accepted XXX. Received YYY; in original form ZZZ}

\pubyear{2015}

\begin{document}
\label{firstpage}
\pagerange{\pageref{firstpage}--\pageref{lastpage}}
\maketitle

\begin{abstract}

We combine new (NGC~1275, NGC~4151, and NGC~5506) and previously published (Cygnus A, Mrk~231, and NGC~1068) sub-arcsecond resolution mid-infrared (MIR; 8-13 \um) imaging- and spectro-polarimetric observations of six Seyfert galaxies using CanariCam on the 10.4-m Gran Telescopio CANARIAS. These observations reveal a diverse set of physical processes responsible for the nuclear polarization, and permit characterization of the origin of the MIR nuclear polarimetric signature of active galactic nuclei (AGN). For all radio quiet objects, we found that the nuclear polarization is low ($<1$ per cent), and the degree of polarization is often a few per cent over extended regions of the host galaxy where we have sensitivity to detect such extended emission (i.e., NGC 1068 and NGC 4151). We suggest that  the higher degree of polarization previously found in lower resolution data arises only on the larger-than-nuclear scales. Only the radio-loud Cygnus A exhibits significant nuclear polarization ($\sim$11 per cent), attributable to synchrotron emission from the pc-scale jet close to the core. We present polarization models that suggest that the MIR nuclear polarization for highly obscured objects arises from a self-absorbed MIR polarized clumpy torus and/or dichroism from the host galaxy, while for unabsorbed cores, MIR polarization arises from dust scattering in the torus and/or surrounding nuclear dust.

\end{abstract}





\section{Introduction}
\label{INTRO}

The mid-infrared (MIR) spectral range has proven to be exceptionally rich in spectral features that can be used to characterize the polarimetric properties of active galactic nuclei (AGN) and their host galaxies. Specifically, the 10 \um~silicate feature can be used to identify the polarization signature of dusty structures, as it has been observed in many young stellar objects, \ion{H}{II} regions, etc.  \cite[i.e.][]{Smith:2000aa}. Also, this spectral range is less affected by obscuration of the host galaxy and/or dust lanes to their core than at optical and near-IR (NIR) wavelengths. However, MIR polarimetry of AGN has been very challenging due to the combination of the relatively low sensitivity of MIR instrumentation to achieve accurate polarimetric measurements of a sample of AGN, and the lack of MIR polarimeters on 8-m class and space-based telescopes. Thus, the origin of the MIR polarization of AGN is still a matter of controversy.

The obscuring torus absorbs a large fraction of the radiation emitted by the AGN, which is then reprocessed and re-emitted in the IR by dust. If non-spherical dust grains are aligned in the torus, then we expect to measure some level of polarization. Furthermore, if the same grains that produce polarization in the NIR are those dominating the MIR emission, then we should expect to measure a MIR polarization signature with a theoretical 90$^{\circ}$ shift of the position angle (PA) of polarization from NIR to MIR wavelengths \citep{Efstathiou:1997aa}. This interpretation is generally consistent with observations \citep{Bailey:1988aa,Brindle:1990ab,Brindle:1990aa,Brindle:1990ac,Young:1996ab,Packham:1997aa,Lumsden:1999aa,Simpson:2002aa,Watanabe:2003aa,Lopez-Rodriguez:2015aa,Gratadour:2015aa} when using moderate (few arcsecond) to high-angular (sub-arcsecond) resolution $1-2$ \um~polarimetric observations of AGN. Using moderate angular resolution MIR imaging- and spectro-polarimetric observations, several studies \citep{Aitken:1984aa,Bailey:1988aa,Lumsden:1999aa,Siebenmorgen:2001aa} found 1) the MIR nuclear polarization of NGC1068 and four Ultra-Luminous IR Galaxies (ULIRGs; including Mrk~231) are in the range of $2-8$ per cent, 2) a 90$^{\circ}$ shift in the PA of polarization from NIR to MIR wavelength, and 3) the spectral shape of the MIR polarized spectrum of NGC 1068 cannot be explained by silicate-type dust grains. These studies have set our current interpretation of MIR polarization of AGN, at scales of few arcseconds, as arising from magnetically aligned dust grains in the dusty torus. The optical to NIR ($0.44-2.2$ \um) polarimetric study of 71 Seyfert galaxies by \citet{Brindle:1990ab,Brindle:1990aa,Brindle:1990ac} represents the largest polarimetric study at moderate angular resolutions to date. Although these authors found that for most objects, dichroism can generally explain the nuclear polarization of Seyfert galaxies, when starlight absorption is taken into account in the polarimetric analysis then dust/electron scattering or a non-thermal component close to the central engine can potentially explain the optical to NIR polarization. However, the large apertures used to measure the polarization of the unresolved cores were the limiting factors in disentangling the dominant nuclear polarization component. Polarimetric studies require multi-wavelength and high-angular resolution observations to disentangle the several competing polarization mechanisms in and around the AGN, and to measure the intrinsic degree of polarization if there are mechanisms producing more than one PA of polarization.

Further sub-arcsecond angular resolution 10 \um~imaging polarimetric observations of the central $\sim200$ pc of NGC 1068 using Michelle on the 8.1-m Gemini North by \citet{Packham:2007aa} challenged the generally accepted physical interpretation that aligned dust grains in the torus is the solely dominant MIR nuclear polarization of AGN. Although the single wavelength observations made it difficult to carry out any further analysis, \citet{Packham:2007aa} suggested to have detected the polarization signature of aligned dust grains in the Northern ionization cones with degree of polarization of a few, and a $<1$ per cent polarized core with an unclear physical interpretation due to the lack of multi-wavelength observations. It was not until recently, with the MIR, 7.5$-$13 \um, imaging- and spectro-polarimeter CanariCam on the 10.4-m Gran Telescopio CANARIAS (GTC) in Spain, that MIR sub-arcsecond ($\sim0.3$ arcsec) polarimetry at relatively high sensitivity has been available to study AGN. The series of papers by \citet{Lopez-Rodriguez:2014aa,Lopez-Rodriguez:2016aa,Lopez-Rodriguez:2017aa} have exploited this capability and show the complexity of the MIR polarimetric properties in the central few hundred pc of nearby and bright AGN. They show, for example: 1) a highly polarized ($\sim11$ per cent) synchrotron nucleus arising from a pc-scale jet close to the core of Cygnus A \citep{Lopez-Rodriguez:2014aa}; 2) a low polarized, $\sim0.3$ per cent, dusty scattered region at a few parsecs from the core of Mrk 231 \citep{Lopez-Rodriguez:2016aa}; and 3) very complex polarized structures within the central $\sim200$ pc of NGC 1068 \citep{Lopez-Rodriguez:2017aa}. These studies also demonstrate significant differences when compared with previously published $1-13$ \um~polarimetric observations. 

In order to provide a general perspective of the MIR polarimetric measurements in the context of the unified model of AGN, we present new, and already published, MIR polarimetric observations of six MIR-bright Seyfert galaxies at sub-arcsecond angular resolution observed with CanariCam on the GTC as part of an ESO/GTC large program  \citep[ID 182.B-2005, PI: Alonso-Herrero,][]{Alonso-Herrero:2016aa}. These MIR polarimetric observations allow us to present a characterization of the origin of the MIR polarization of AGN at sub-arcsecond scales. The paper is organized as follows: Section \ref{OBS} describes the observations and data reduction, Section \ref{RES} shows the results of the new observed AGN and the analysis of their nuclear polarization. Section \ref{Dis} discusses the origin of the MIR polarization for our sample of AGN. In Section \ref{CON} we present our conclusions.



\section{Observations and Data Reduction}
\label{OBS}

Imaging- and spectro-polarimetric modes \citep{Packham:2005aa} of CanariCam \citep{Telesco:2003aa} on the GTC were used to perform observations of a MIR-bright sample of AGN (Table \ref{table1}), limited by the sensitivity of the polarimetric modes of CanariCam. All data were observed in queue mode under photometric conditions and image quality better than 0.6 arcsec measured at MIR wavelengths from the full width half maximum (FWHM) of the standard stars.

CanariCam uses a 320 $\times$ 240 pixel Si:As Raytheon array, with a pixel scale of 0.0798 arcsec pixel$^{-1}$. The polarimetric mode uses a half-wave retarder (half-wave plate, HWP), a field mask, and a Wollaston prism, where a slit and grating are introduced in the optical path to obtain spectro-polarimetric observations. The Wollaston prism and HWP are made with sulphur-free CdSe. The HWP is chromatic, resulting in a variable polarimetric efficiency across the 7.5$-$13 \um~wavelength range. In all observations, the HWP is set to four PA, in the following sequence 0$^{\circ}$, 45$^{\circ}$, 22.5$^{\circ}$ and 67.5$^{\circ}$, and the central polarization slot with a FOV of 25.6 arcsec $\times$ 2.0 arcsec was used. We use a standard chop-nod technique to remove time-variable sky background and telescope thermal emission, and to reduce the effect of 1/\textit{f} noise from the array/electronics.


\begin{table}
	\caption{Details of the new observed Seyfert galaxies.}
	\label{table1}
	\begin{tabular}{lcccccccc}
		\hline
Object	&	Redshift	&	D$^{\mbox\scriptsize{{(a)}}}$ 	&	Scale 		&	Type	\\
		&			& (Mpc)	&	 (pc/arcsec)	&	 \\
\hline
NGC~1275	&	0.017559	&	70.9	&	332	&	RG/Sy1.5	\\
NGC~4151	&	0.003319	&	17.1	&	82	&	Sy1.5	\\
NGC~5506	&	0.006181	&	29.1	&	139	&	Sy1	\\
\hline
	\end{tabular} \\
	$^{\mbox\scriptsize{{(a)}}}$Distances estimated adopting H$_{0}$ = 73 km s$^{-1}$ Mpc$^{-1}$
\end{table}



\begin{table*}
	\caption{Summary of observations.}
	\label{table2}
	\begin{tabular}{lcccccccc}
\hline
Object	&	Date			&	Obs. Mode	&	Filters	&	Chop-angle 	&	IPA$^{(1)}$			&	On-source time		&	FWHM$_{\mbox{\scriptsize{PSF}}}^{(2)}$\\
		&	(yyyymmdd)	&				&			&	($^{\circ}$)	&	($^{\circ}$)	&		(s)			&	(arcsec)\\
\hline
NGC~1275	&	20150916				&	Spec-pol	&	N-band			&	90	&	0	&	1852					& 0.30$-$0.35\\
NGC~4151	&	20150311				&	Ima-pol	&	Si2, Si4, Si5, PAH2	&	30	&	60	&	1748, 1748, 1904, 1894	& 0.40$-$0.45\\
				&	20150326, 20150403	&	Spec-pol	&	N-band			&	170	&	80	&	5626					& 0.45\\
NGC~5506	&	20150404				&	Ima-pol	&	Si2, Si4, Si6		&	90	&	0	&	728, 728, 888			& 0.35$-$0.45\\
\hline
	\end{tabular} \\
$^{(1)}$Instrumental Position Angle (IPA), the angle of the short axis of the array with respect to the North on the sky.  $^{(2)}$FWHM of the PSF.
\end{table*}


In the imaging-polarimetric mode, the Si2 ($\lambda_{\mbox{\scriptsize c}}$ = 8.7 \um, $\Delta\lambda$ = 1.1 \um, 50 per cent cut-on/off), Si4 ($\lambda_{\mbox{\scriptsize c}}$ = 10.3 \um, $\Delta\lambda$ = 0.9 \um, 50 per cent cut-on/off) and Si5 ($\lambda_{\mbox{\scriptsize c}}$ = 11.6 \um, $\Delta\lambda$ = 0.9 \um, 50 per cent cut-on/off) filters provide the largest wavelength coverage with the best combination of sensitivity and spatial resolution for the filter set of CanariCam in the 10 \um~atmospheric window. Additionally, we performed observations using the PAH2 ($\lambda_{\mbox{\scriptsize c}}$ = 11.3 \um, $\Delta\lambda$ = 0.6 \um, 50 per cent cut-on/off) filter. In the spectro-polarimetric mode, a 0.41 arcsec ($\sim$5 pixels) wide slit was used. The low spectral resolution N-band ($\lambda_{\mbox{\scriptsize c}}$ = 10.4 \um, $\Delta\lambda$ = 5.2 \um, 50 per cent cut-on/off) grating was used, resulting in a dispersion of 0.019 \um~pixel$^{-1}$ and R =$\lambda$/$\Delta\lambda \sim$ 175. In all observations, the chop-throw was 8 arcsec. Table \ref{table2} summarizes the observational details.

Data were reduced using custom \textsc{idl} and \textsc{python} routines. For all observations, the difference for each chopped pair was calculated and the nod frames then differenced and combined to create a single image per HWP PA. During this process, all nods were examined for high or variable background that could indicate the presence of clouds or variable precipitable water vapour. Fortunately, no data needed to be removed for these reasons. As the galaxies were observed in different sets, each HWP PA frame was registered and shifted to a common position, then images with the same HWP PA were co-added. Next, the ordinary (o-ray) and extraordinary (e-ray) rays, produced by a Wollaston prism, were extracted and the Stokes parameters I, Q and U were estimated according to the ratio method \citep[e.g.][]{Tinbergen:2005aa}. Then, the degree $P = \sqrt{Q^2 + U^2}/I$ and $PA = 0.5 \arctan{(U/Q)}$ of polarization were estimated. The instrumental polarization and polarization efficiency were corrected as described by \cite{Lopez-Rodriguez:2014aa}. Bright, $> 80$ Jy, and polarized, $\sim1-3$\%, young stellar objects, i.e. AFGL 490, AFGL 989 and AFGL 2403, were observed to estimate the zero-angle calibration. The measurements of the degree of polarization were corrected for polarization bias using the approach by \citet{Wardle:1974aa}.

For each spectropolarimetric observational set, the o- and e-rays were extracted using a point-spread function (PSF)-extraction at the position of the peak pixel in total flux. CanariCam shows a slight ($\sim$1 pixel) curvature across the array in the spectral direction, which was measured using the observations of the associated polarized young stellar object to each observation. This curvature was taken into account during the extraction of the spectra. Although the variation of the polarization efficiency was accounted for, a variable signal-to-noise ratio (SNR) of the Stokes parameters as a function of the wavelength still remains. We corrected this effect by binning the Q and U spectra across the wavelength axis to obtain a polarization signal of $P/\sigma_{P} >$ 3 for each bin, where $\sigma_{P}$ is the uncertainty of the polarization measurement of each bin.


\section{Polarimetric Analysis of individual objects}
\label{RES}

\subsection{NGC 1275}
\label{Ana:NGC1275}

Due to scheduling constraints, not all the requested datasets were obtained resulting in a low SNR polarimetric spectrum. Despite the limited observations, an upper limit of the degree of polarization of 0.4 per cent within the 8-12 \um~wavelength range is estimated (Table \ref{table3}) while the PA  of the polarization could not be constrained.

Although this object has been broadly studied at all wavelengths, sub-arcsecond angular resolution polarimetric observations at optical and IR wavelengths are lacking. \citet{Kemp:1977aa} observed a polarization of 0.52 $\pm$ 0.20 per cent with a PA of polarization of 56 $\pm$ 15$^{\circ}$ in a 5.9 arcsec (1.9 kpc) beam at 2.2 \um~using photoelastic modulators on the 2.25-m Steward Observatory telescope. \citet{Rudy:1982aa} reported a statistically insignificant polarization detection at 2.3 \um~in a 7.5 arcsec (2.5 kpc) beam using the UC San Diego (UCSD) polarimeter on the 3-m Shane Telescope at Lick Observatory. Our 8$-$13 \um~observations show an upper-limit polarization of 0.4 per cent in a 0.4 arcsec (133 pc) aperture. The degree of polarization is invariant to the aperture of the observations, which indicates that the dominant polarization arises from the unresolved core. We tentatively interpret that the low polarization arises from dilution by interstellar polarization of the host galaxy and by our Galaxy (NGC 1275 has a low galactic latitude). The poor angular resolution of these previous studies and the lack of a multi-wavelength polarimetric observations make any further interpretation of the nuclear polarization of NGC 1275 difficult.

\subsection{NGC 4151}
\label{Ana:NGC4151}
 
\begin{figure*}
\includegraphics[angle=0,trim=0cm 1cm 0cm 0cm,scale=0.34]{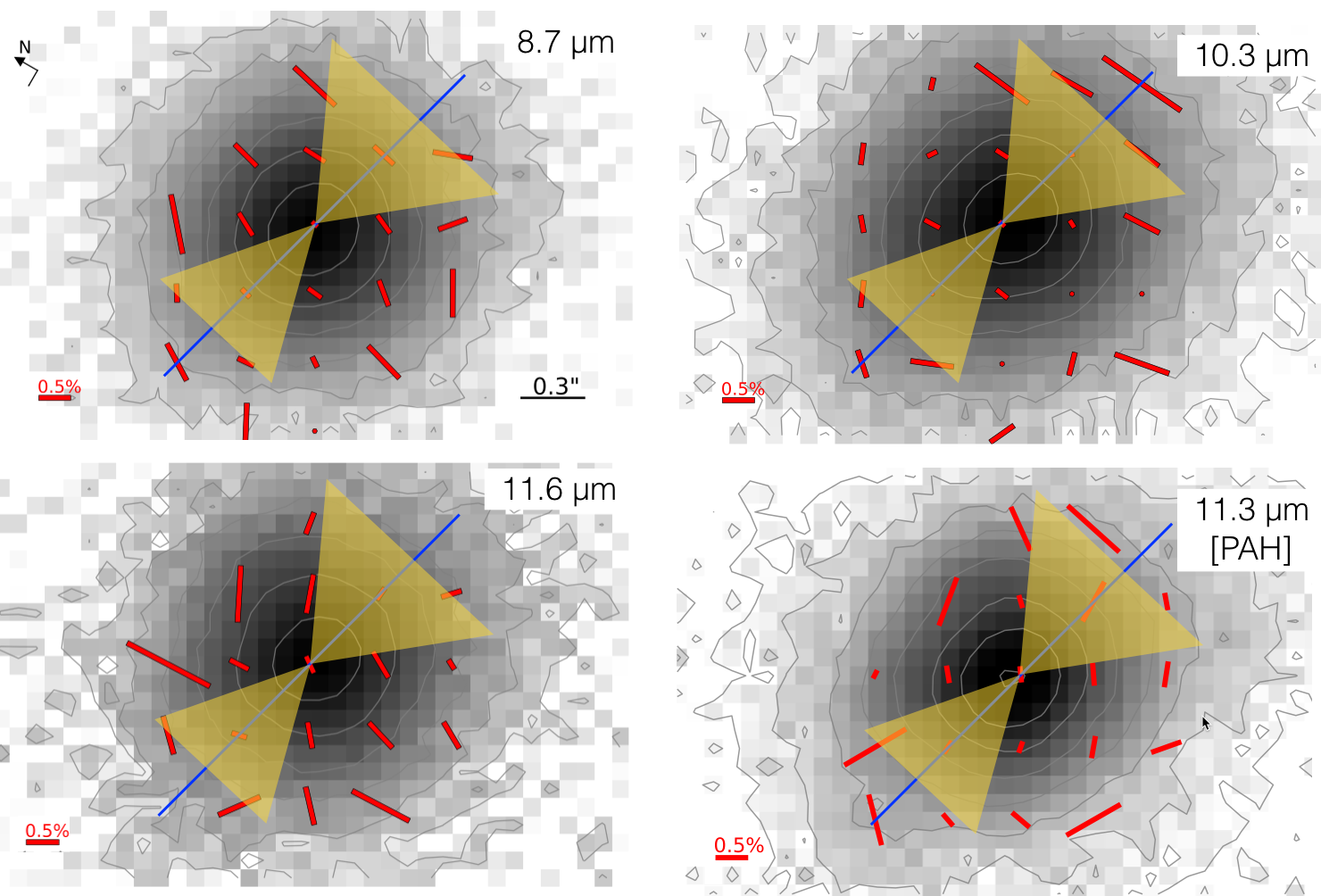}
\caption{NGC 4151: Imaging-polarimetric observations of the central 3 $\times$ 2 arcsec (246 $\times$ 164 pc) at 8.7 \um~(top-left), 10.3 \um~(top-right), 11.6 \um~(bottom-left), and 11.3 \um~(bottom-right). The total flux is shown in grayscale and contours increase as $2^{n}\sigma$, where $n = 2, 3, 4, \dots$. Each polarization vector (red) is statistically independent. Only vectors with $P/\sigma_{P} > 3$ are shown. The radio jet axis (solid blue line) and opening angle of the ionization cones (yellow shaded areas; \citet{Das:2005aa}) are shown. A 0.5 per cent polarization vector is shown at the bottom left of each panel. North and East are shown at the top-left figure.}
\label{fig1}
\end{figure*}

\begin{figure}
\includegraphics[angle=0,trim=0cm 0.5cm 0cm 0cm,scale=0.65]{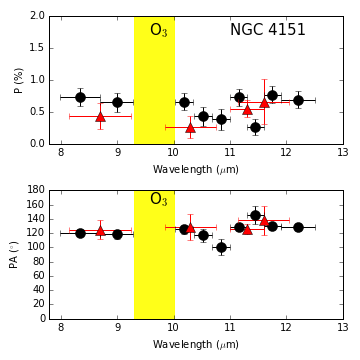}
\caption{NGC 4151: degree (top) and PA (bottom) of polarization of the nuclear region. Imaging- (red triangles) and spectro-polarimetric (black circles) measurements are shown. The O$_{3}$ atmosphere band (yellow shaded region) is shown. }
\label{fig2}
\end{figure}

Figure \ref{fig1} shows the total flux images at 8.7 \um~(top-left), 10.3 \um~(top-right), 11.6 \um~(bottom-left), and 11.3 \um~(bottom-right) of the 3 $\times$ 2 arcsec (246 $\times$ 164 pc) central regions of NGC 4151. The overlaid polarization vectors are proportional in length to the degree of polarization, and their orientation shows the PA of polarization per bin, 0.4 arcsec. Only polarization vectors with $P/\sigma_{P} > 3$ are shown. Figure \ref{fig2} shows the degree and PA of polarization of the spectro-polarimetric observations. Imaging- and spectro-polarimetric observations are consistent with a core with a median degree of polarization of $\sim$0.6 per cent and PA of polarization of $\sim$125$^{\circ}$ in a 0.4 arcsec~( 33 pc) aperture within the 8$-$12 \um~wavelength range (Table \ref{table3}). 

The total flux of NGC 4151 shows a resolved nucleus elongated at a PA of 60$^{\circ}$ East of North spatially coincident with the narrow line region (NLR), in excellent agreement with previous N-band imaging using OSCIR/Gemini-North by \citet{Radomski:2003aa}. The PA of polarization within the ionization cones is nearly perpendicular to the radio-jet axis at all wavelengths. The South-West ionization cone at 11.6~\um~and 11.3 \um~are the exception, with a 90$^{\circ}$ shift in the PA of polarization with respect to the rest of the observations. This result may suggest a change in the polarization mechanism, i.e. from dichroic absorption to dichroic emission, however a 1-13 \um~multi-wavelength polarimetric study of this extended polarized region will be necessary to obtain further conclusions.


NGC 4151 is a type 1.5 AGN which implies that we have a direct view of the central engine with the postulated torus in a near face-on view. However, the  inclination of the AGN has been estimated \citep[see][for a compilation]{Marin:2016aa}  to be in the range of $9-63^{\circ}$ depending on the methodology used, i.e. X-ray spectra fitting using disk dispersion models, IR spectra fitting using clumpy torus models, BH mass and velocity dispersion relationship, etc.  indicating the complexity and potentially different structures in the central few parsecs of this galaxy. In addition, the radio-jet has been observed with an averaged PA $\sim77^{\circ}$ using arcsecond angular resolution observations, however sub-arcsecond resolution observations \citep{2003ApJ...583..192M} have found that the inner few pc of the radio-jet is aligned with a PA $\sim90^{\circ}$. This study shows a potential misalignment between small and large components in the radio-jet. Thus, as we are analyzing sub-arsecond angular resolution MIR polarimetric observations, a PA $\sim90^{\circ}$ \citep{2003ApJ...583..192M} for the radio-jet is the most suitable for the discussion of the MIR nuclear polarization.  An example of this kind of misalignment is found in the central few tens of parsec of NGC 1068. Specifically, \citet{Lopez-Rodriguez:2016aa} found a change in the degree and PA of MIR polarization where the jet changes direction when hits a giant molecular cloud at $\sim0.4$ arcsec (24 pc) North of the core. The radio-jet axes of the central few parsec is required to interpret the MIR polarimetric observations.

NGC 4151 was observed by \citet{Ruiz:2003aa} at J (1.23 \um) and H (1.64 \um) bands using the dual-beam polarimeter IRPOL2 on the 3.8-m United Kingdom Infrared Telescope (UKIRT). They reported a polarization of $P_{J} = 0.78 \pm 0.08 $ per cent and $P_{H} = 0.56 \pm 0.06 $ with $PA_{J} = 116.7 \pm 2.9^{\circ}$ and $PA_{H} = 122.7 \pm 3.1^{\circ}$, respectively in a 1 arcsec (82 pc) aperture, which is similar to our 8$-$13 \um~wavelength nuclear polarization (Table \ref{table3}). The nearly constant degree and PA of polarization from 1 \um~to 12 \um~strongly suggest that a single polarization mechanism dominates in the core of NGC 4151.

Scattering off optically thick dust close to the central engine will produce approximately constant degree and PA of polarization as a function of wavelength (i.e. grey scattering). This polarization mechanism is consistent with the observed nearly constant degree and PA of polarization from 1 \um~to 12 \um. Other polarization mechanisms, i.e. synchrotron emission and/or dichroism, can be ruled out because we do not observe the expected wavelength dependence of the degree and PA of polarization and polarized flux from them. To obtain low polarization by scattering, the physical component scattering off radiation must be a nearly symmetrical distribution of matter to cancel out the centrosymmetric polarization pattern around the central engine. This symmetrical distribution of dust must be within our aperture of 0.4 arcsec (33 pc). Thus, the most likely dominant polarization in the unresolved core of NGC~4151 arises from dust scattering from optically thick dust grains from a symmetrical distribution of dust surrounding the central engine, for example a nearly face-on (inclination $\le45^{\circ}$) torus or dust at the base of the ionization cone.

\subsection{NGC 5506}
\label{Ana:NGC5506}

At all wavelengths, a point-like source is observed with variations at the lowest-surface brightness due to PSF variations. An upper-limit of the degree of polarization of 0.5 per cent in a 0.4 arcsec aperture is measured in the 8$-$12 \um~wavelength range, while the PA  of polarization could not be constrained (Table \ref{table3}).

NGC 5506 is a type 1 AGN with a nucleus highly extinguished by the host galaxy. Recent optical spectro-polarimetric observations \citep{Ramos-Almeida:2016aa} using the FOcal Reducer and low dispersion Spectrograph 2 (FORS2) on the 8 m Very Large Telescope (VLT) reported an intrinsic (starlight corrected) degree of polarization of the continuum emission in the V band of $11.5 \pm 0.2$ per cent with a PA of polarization of $75 \pm 2^{\circ}$ in a 0.9 arcsec (125 pc) aperture. As shown by \citet{Ramos-Almeida:2016aa}, the measured nuclear polarization is highly diluted by the host galaxy (A$_{\mbox{v}} \ge 11$ mag; \citet{1994ApJ...422..521G}), with an estimated host galaxy contribution of 81\% at 0.55 \um. Hidden broad polarized emission lines in the optical polarized spectra were not found by \citet{Ramos-Almeida:2016aa}, however the NLR profiles became broader in the IR \citep{1994ApJ...422..521G}. \cite{2013ApJS..209....1F} observed one side of the ionization cone with a $\sim$5 arcsec extension and PA $\sim 22^{\circ}$ using [\ion{O}{iii}] \textit{HST} imaging observations. As the IR is less sensitive to extinction than optical wavelengths, the NLR broad profiles from the less attenuated ionization cone become easier to detect. These results suggest that the nucleus is highly extinguished across the core, with the Northern ionization cone inclined slightly towards our LOS, avoiding the high extinction by the host galaxy. 

\begin{figure}
\includegraphics[angle=0,trim=0.8cm 1cm 0cm 0cm,scale=0.43]{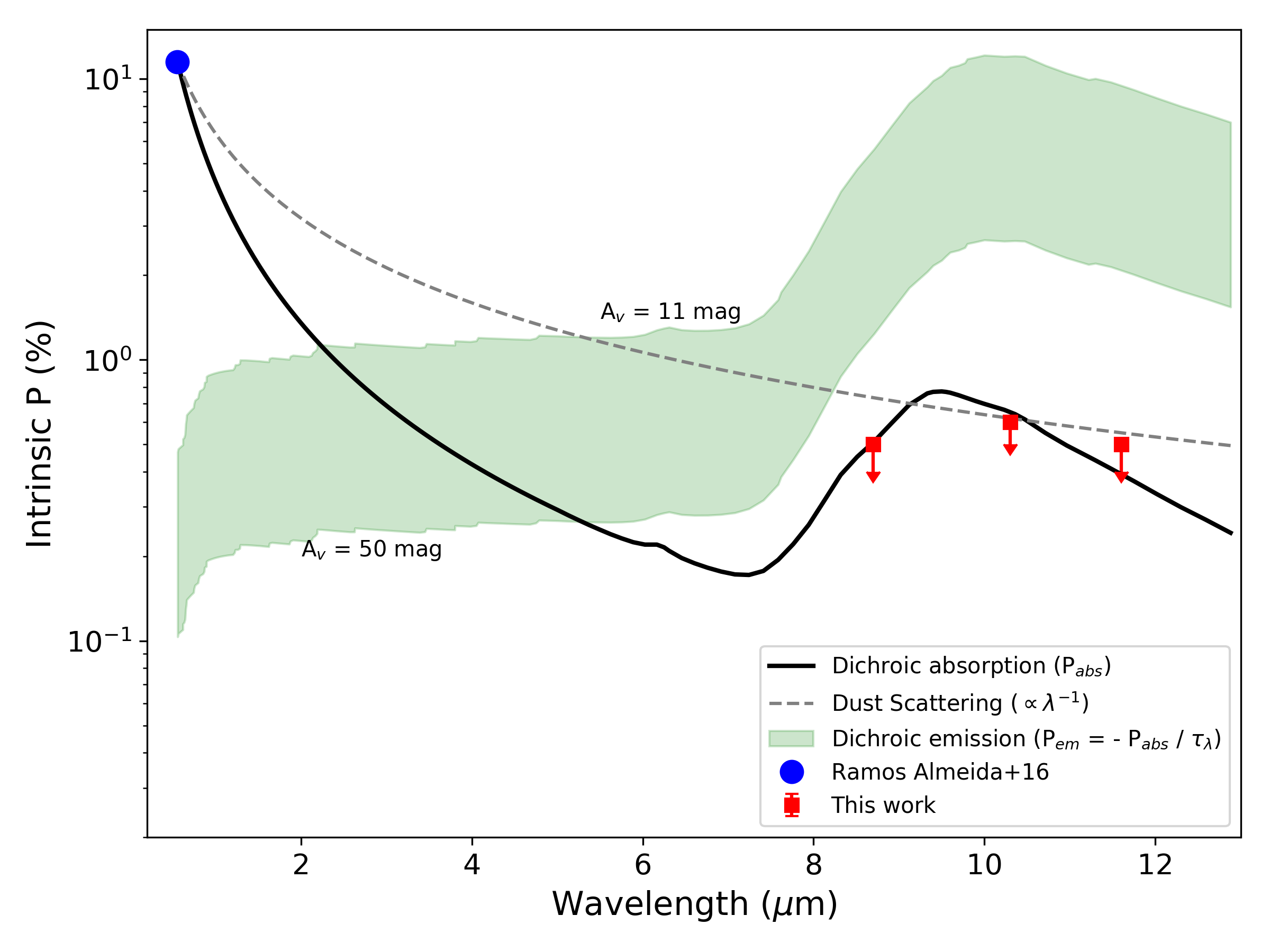}
\caption{Expected intrinsic polarization of NGC 5506 at MIR wavelengths by dichroic absorption (black solid line) and emission (green shadowed area) from aligned dust grains, and dust scattering, $P \propto \lambda^{-1}$ (grey dashed line). Dichroic absorption and dust scattering models were normalized to the observed polarization (blue circle) of 11.5 $\pm$ 0.2 per cent at 0.55 \um~\citep{Ramos-Almeida:2016aa}. CanariCam observations (red squares) are shown.}
\label{fig3}
\end{figure}

We investigate if the MIR polarization arises from high obscuration of the nuclear emission by the host galaxy, i.e. dichroic absorption. \cite{Roche:2007aa} suggested that the extinction measured from the 9.7 \um~silicate feature arises from the interstellar medium of the host galaxy of NGC 5506. If the extinction is originated in the torus, we would detect a shallower 9.7 \um~silicate feature. To account for the absorptive polarization component from optical to MIR wavelengths, we use the V-band polarization measurement as the normalization of the wavelength-dependence of the dichroic absorption component \citep{Serkowski:1975aa} up to 5 \um~and assume typical silicate dust grains at MIR wavelengths. This approach is similar to that followed by \cite{Lopez-Rodriguez:2016aa} to explain the polarization from dichroism in the core of NGC 1068. Figure \ref{fig3} shows the expected absorptive (black solid line) and emissive (green shadowed area) polarization components, which shows that an absorptive polarization of $\sim$0.6 per cent is expected in the 8$-$12 \um~wavelength range. This result is consistent with our measured upper-limit of 0.6 per cent in the same wavelength range, which suggests that the observed polarization of NGC 5506 in the 8$-$12 \um~wavelength range arises from dichroic absorption by aligned dust grains in the host galaxy. This result suggests that the dust component that dilutes the emission from the central engine into our LOS is the same as the component observed in total flux containing a large concentration of silicates. We note that the absorptive component presents multiple degrees of degeneracy, i.e. dust composition, extinction, etc., that can be used to scale the silicate feature. We here have shown that a standard silicate feature normalized to the optical intrinsic polarization can reproduce acceptedly well our MIR polarimetric measurements.

We test other alternative polarization mechanisms: 1) dust emission from aligned dust grains in the torus, and 2) dust scattering from the wall of the torus. For the former, Fig. \ref{fig3} shows that if we have a direct view of dust emission from the torus, i.e. dichroic emission (green shadowed area), then an intrinsic emissive polarization of $\ge 5$ per cent in the $8-12$ \um~wavelength range should be measured, which is inconsistent with our observations. For the latter, Fig. \ref{fig3} shows the expected dust scattering polarization in the form of $\propto \lambda^{-1}$ above the observed upper-limits from our observations. Thus, we can rule out polarization arising from dust emission by aligned dust grains and/or dust scattering in the torus of NGC 5506.


\begin{table}
\caption{Nuclear polarization.}
\label{table3}
\begin{tabular}{ccccccc}
\hline
Object	&	Aperture	&	$\lambda$		&	P	&	PA 	\\
		&	(arcsec)		&	(\um)		&	 (\%)	&	($^{\circ}$) \\
\hline
NGC~1275	&	0.40	&	8-12$^{\mbox{\scriptsize{a}}}$	&	$<$0.4	&	-	\\
NGC~4151	&	0.40	&	8.7	&	0.4 $\pm$ 0.2	&	125 $\pm$ 13		\\
			&		&	10.3	&	0.3 $\pm$ 0.2	&	129 $\pm$ 18	\\
			&		&	11.3	&	0.6 $\pm$ 0.1	&	126 $\pm$ 7	\\
			&		&	11.6	&	0.7 $\pm$ 0.4	&	138 $\pm$ 20	\\
			&		&	8-12$^{\mbox{\scriptsize{a}}}$	&	0.5 $\pm$ 0.2	&	125 $\pm$ 8	\\
NGC~5506	&	0.40	&	8.7	&	$<$0.5	&	-	\\
			&		&	10.3	&	$<$0.6	&	-	\\
			&		&	11.6	&	$<$0.5	&	-	\\
\hline
\end{tabular} \\
$^{\mbox{\scriptsize{a}}}$Averaged polarization within the 8$-$12 \um~wavelength range using our spectro-polarimetric observations. \\
\end{table}


\begin{table*}
\caption{Dominant polarization mechanisms and physical components in the central regions of the AGN sample.}
\label{table4}
\begin{tabular}{ccccccp{6cm}}
\hline
Object		&	Type			&	Aperture	&	Aperture	&	DPA$^{\mbox{\scriptsize{a}}}$			&	Polarization Mechanism	&	Physical component \\
			&				&	(arcsec)	&	(pc)		&	($^{\circ}$)	\\
\hline
Cygnus~A		&	RG/FRII		&	0.38		&	392		&	75		&	Synchrotron emission	& pc-scale jet close to the core\\
Mrk~231		&	ULIRG/Sy1	&	0.50		&	404		 &	40		&	Dust scattering			& hot dust in the pc-scale dusty regions\\
NGC~1068	&	Sy2			&	0.38		&	 29		&	--		&	Dichroic emission$^{\mbox{\scriptsize{b}}}$		& torus\\
NGC~1275	&	RG/Sy1.5		&	0.40		&	133		&	--		&	Unclear$^{\mbox{\scriptsize{c}}}$ 				& Unclear \\
NGC~4151	&	Sy1			&	0.40		&	33		&	35		&	Dust scattering			& torus/base of ionization cone \\
NGC~5506	&	Sy2			&	0.40		&	56		&	--		&	Dichroic absorption		& host galaxy\\
\hline
\end{tabular} \\
$^{\mbox{\scriptsize{a}}}$Difference between the PA of polarization at MIR wavelength with the PA of the radio-jet. Radio-jet angles: Cygnus A \citep{Sorathia:1996aa}, Mrk231 \citep{Reynolds:2017aa}, NGC 4151 \citep{Pedlar:1998aa}\\
$^{\mbox{\scriptsize{b}}}$This result is based on the upper-limit  polarization measurements and modeling by \citet{Lopez-Rodriguez:2016aa}. \\
$^{\mbox{\scriptsize{c}}}$Not enough polarimetric observations to disentangle the dominant polarization mechanism.
\end{table*}


\section{MIR nuclear polarization}
\label{Dis}


\begin{figure*}
\includegraphics[angle=0,trim=0cm 0.5cm 0cm 0cm,scale=0.55]{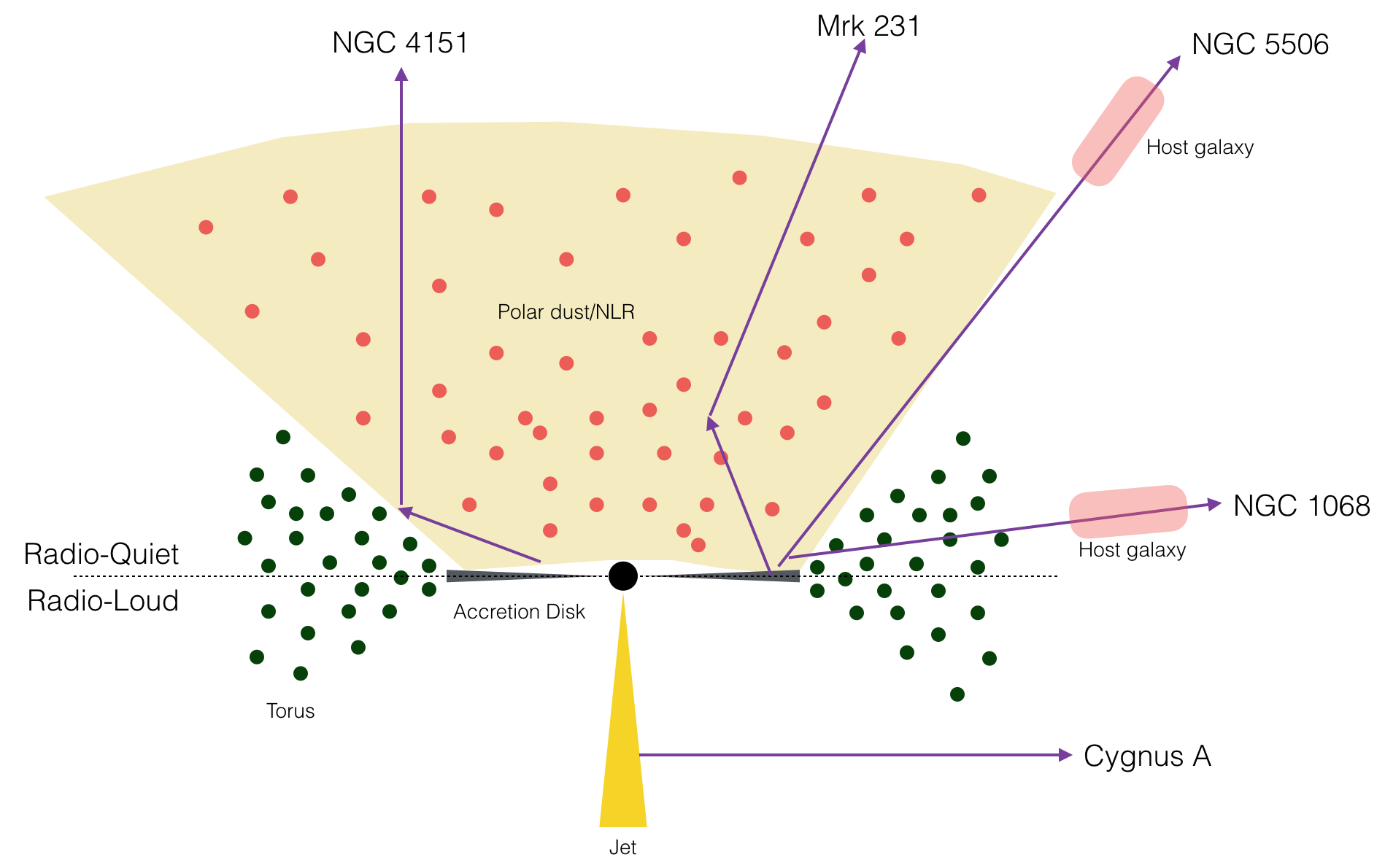}
\caption{Sketch of the central few pc of a radio-quiet (Radio-quiet, RQ; top panel) and radio-loud  (Radio-loud, RL; bottom-panel) AGN. The expected radiation path of the observed polarization at MIR wavelength for each object in this work is shown. Section \ref{Dis} for details.}
\label{fig4}
\end{figure*}


We summarize our sub-arcsecond angular resolution MIR nuclear polarization observations in Table \ref{table4}, and Figure \ref{fig4} shows a sketch of the physical interpretation of the path of the observed dominant polarization of each object. Specifically, for the radio-quiet objects with an nearly edge-on view and heavily extinguished (NGC 1068, NGC 1275, and NGC 5506), the nuclear polarization arises from dichroism. NGC 1068 nuclear polarization arises from a self-absorbed dichroic emission in the dusty torus surrounding the central engine. NGC 1275 nuclear polarization likely arises from dichroic absorption by aligned dust grains in the galaxy. NGC 5506 nuclear polarization arises from dichroic absorption of silicate grains in the galaxy. For the radio-quiet objects with a nearly face-on view and unobscured cores, (Mrk 231 and NGC 4151), the polarization arises from dust scattering. Mrk 231 polarization arises from scattering off dust from a symmetrical distribution of hot dust within the central $\sim2$ pc around its core. NGC 4151 nuclear polarization arises from dust scattering from optically thick dust grains from a symmetrical distribution of dust surrounding the central engine, for example a nearly face-on (inclination $\le45^{\circ}$) torus or dust at the base of the ionization cone.  For the only radio-loud AGN, Cygnus A nuclear polarization arises from synchrotron emission from the pc-scale jet close to the core. 

To sum up, the nuclear MIR polarization is low ($<$ 0.4 per cent) and the MIR polarization arises from different physical components and polarization mechanisms in and around the AGN. These observational results challenge the general interpretation that the MIR polarization in the core of AGN arises from aligned dust grains in the torus. Thus, the diversity of physical mechanisms found in these AGN shows that there is no one mechanism dominating the MIR nuclear polarization of AGN. Instead, several polarization mechanism are competing at the MIR wavelengths and it seems necessary to model individual objects to obtain as much physical information we can from each one of them. 

Although the degree of polarization in our radio-quiet sample is always low ($<1$ per cent) over the nucleus, there where is enough SNR to detect extended polarization over the host galaxy (i.e. NGC 1068 and NGC 4151), the degree of polarization is often high. For  NGC 1068, the extended polarization peaks at $\sim 7$ per cent due to absorption of polarized emission from the collision between the jet and the giant molecular cloud in the Northern ionization cone at $\sim 0.4$ arcsec (24 pc) from the core. For NGC 4151, the extended polarization peaks at $\sim 3$ per cent within the central $\sim$1.5 arcsec ($\sim123$ pc) along the ionization cones. We find that the interpretation by \citet{Siebenmorgen:2001aa}, that the AGN torus is highly polarized in the MIR, is likely dominated by the host galaxy and not the AGN. Further sensitive and deeper polarimetric observations are required to characterize the dust characteristics of the extended dust emission around the AGN.

\subsection{MIR polarization of the dusty torus}

We have argued in \citet{Lopez-Rodriguez:2016aa} for the Type 2 AGN in NGC 1068 that its low polarized ($<0.4$ per cent) core arises from dichroic emission from aligned dust grains. Indeed, the only models \citep{Efstathiou:1997aa} that show any net degree of polarization, assume a homogeneous torus with polarization levels in the range of few per cent in the MIR for edge-on AGN and consistent with null polarization for face-on AGN. This is explained as the outer layers of the homogeneous torus, without dilution with the obscuring by dusty medium, will be directly observed enhancing the net emissive polarization. However, these models fail to reproduce the current sub-arcsecond angular resolution MIR polarimetric observations. If a clumpy distribution of dust in the torus is assumed, we have suggested that although the obscuring material may be intrinsically polarized at NIR wavelengths, the MIR polarized emission is self-absorbed by the dusty medium within the clumpy distribution, causing a net emissive polarized component to be diluted once the radiation leaves the dusty environment. 

Recent modeling efforts have brought new insights into the dynamics of the torus. Specifically, the magnetohydrodynamical (MHD) simulations by \citet*{2017ApJ...843...58C} have shown that a mid-plane inflow with a high-latitude outflow launched from the torus inner edge by UV radiation and then driven by IR radiation best describes the torus. In addition, they found that although the torus is magnetized, the wind is driven by radiation pressure with density inhomogeneities imprinted into the wind that can be seen as the current scheme of a `clumpy' torus. Recent MIR interferometric observations \citep[e.g.][]{Honig:2012aa,Honig:2013aa,Lopez-Gonzaga:2016aa} observed that up to $60-90$ per cent of the MIR nuclear emission in Seyfert galaxies could be associated with a polar-elongated dusty region within the few pc of the AGN. It is worth mentioning that for environments where dust is directly heated by a strong radiation field, i.e. asymptotic giant branch C-rich stars, young stellar objects or AGN, radiation pressure can align the dust grains more efficiently than magnetic alignment and produce some measurable degree of polarization. This is explained by the currently accepted radiative torque dust grain alignment \citep[RATs,][]{Hoang:2008aa}.  \citet{2017arXiv171202192G} used a radiative transfer code, MontAGN, optimized for IR polarimetric observations and assume spherical dust grains for a region with lower optical depth and larger scales than the torus. Although these models have been able to reproduce large-scale scattering patterns when compared with the 1.6 \um~and 2.2 \um~imaging polarimetric observations with SPHERE/VLT by \citet{Gratadour:2015aa}, the nuclear polarization is still unclear. Thus, further MHD simulations with polarimetric capabilities will be crucial to shed light to the high-angular resolution MIR observations presented here. If our model of the MIR polarization being self-absorbed emission is incorrect, the fact that the MIR polarization shows low polarization would be inconsistent with dichroism in the torus of NGC 1068. If so, this would imply that the NIR to MIR polarization detected by many is instead due to scattering within the resolution limit of those observations and not to dichroism. 

In fact, for Mrk 231 and NGC 4151, the dust distribution and inclinations of the physical components of AGN are such that dust scattering can explain the measured $1-13$ \um~ nuclear polarization. Specifically, dust in the torus and/or in the pc-scale central regions explain acceptably well their polarized IR SED. These results can potentially imply that the scattering efficiency and sizes of dust grains in the MIR are larger in the close environments of AGN. This physical properties have been observed in dense molecular clouds by \citet{2016A&A...585L...4L}, where non-negligible scattering efficiency at MIR wavelengths have been measured, and suggested in AGN \citep{2001A&A...365...37M}. High-angular resolution $1-13$ \um~polarimetric observations with sensitive polarization at large scales, as those by \citet{Lopez-Rodriguez:2015aa,Gratadour:2015aa,Lopez-Rodriguez:2016aa}, in a large AGN sample together with developed models \citep{1996ApJ...468..606K,1999ApJ...518..676K} are required to test the hypothesis of scattering as the dominant polarization mechanisms of AGN.


\section{Conclusions}
\label{CON}

We presented new 8$-$12 \um~imaging- and spectro-polarimetric sub-arcecond angular resolution observations using CanariCam on the 10.4-m GTC of three Seyfert galaxies NGC 1275, NGC 4151 and NGC 5506. For these three galaxies, we found a low polarization ($<0.5$ per cent) core. For NGC 4151, the nuclear polarization most likely arises from dust scattering from a symmetrical distribution of dust surrounding the central engine (i.e. torus or dust at the base of the ionization cone). We found an extended polarized emission along the ionization cones with degrees of polarization of few percent and PA of polarization nearly perpendicular to the radio axis. For NGC 5506, the nuclear polarization most likely arises from dichroic absorption by aligned dust grains in the host galaxy. For NGC 1275, the lack of high-angular resolution multi-wavelength polarimetric observations makes any interpretation of the nuclear polarization difficult.

We compiled the MIR high-angular resolution polarization publicly available observations of AGN (nearby and bright Seyfert galaxies), and found that the degree of polarization in this radio-quiet sample is always low ($< 1$\%) over the nucleus. Only the radio-loud Cygnus A exhibits high polarization ($\sim$11 per cent) in the core. Although the most straight forward result is the MIR polarization arising from synchrotron emission by the pc-scale jet close to the core of Cygnus A, the reality is that AGN show a very complex MIR nuclear polarization highly dependent on their geometry, morphology and physical components. 

For NGC 1068 and NGC 4151, we had enough SNR to detect extended polarization spatially coincident with the ionization cones and with high degree of polarization ($2-7$ per cent). We suggest that by using the sub-arcsecond resolution polarimetric observations, we find that the interpretation by \citet{Siebenmorgen:2001aa}, that the AGN torus is highly polarized, is likely dominated by the host galaxy and not the AGN. 

We here reported an update of the current observations and knowledge of MIR polarimetry of AGN using sub-arcsecond observations. However, it is necessary to perform deeper observations to obtain better polarimetric sensitivity, and larger sample to perform a statistical analysis based on AGN types, luminosities, morphologies, etc.

\vspace{-0.5cm}

\section*{Acknowledgments}

The referee is especially thanked for a thorough reading of the original submitted paper, and making numerous useful comments to improve it. Based on observations made with the Gran Telescopio CANARIAS (GTC), installed in the Spanish Observatorio del Roque de los Muchachos of the Instituto de Astrof\'isica de Canarias, in the island of La Palma. A.A.H. acknowledges financial support from the Spanish Ministry of Economy and Competitiveness through the Plan Nacional de Astronom\'ia y Astrof\'isica under grant AYA2015-64346-C2-1-P , which is partly funded by the FEDER programme. O.G.M. acknowledges financial support from the UNAM PAPIIT IA100516 and IA103118. C.R.A. acknowledges the Ram\'on y Cajal Program of the Spanish Ministry of Economy and Competitiveness through project RYC-2014-15779 and the Spanish Plan Nacional de Astronom\' ia y Astrofis\' ica under grant AYA2016-76682-C3-2-P. K.I. acknowledges the financial support by the Grant-in-Aid for JSPS fellow for young researchers (PD). C.P. acknowledges financial support from NSF-1616828. Analysis done with \textsc{astropy}\footnote{\url{http://www.astropy.org/}} routines.

\bibliographystyle{mnras}
\bibliography{Atlas_Ref}


\bsp	
\label{lastpage}
\end{document}